\documentstyle[12pt,epsf]{article}
\begin{document}
\vspace*{2cm}


\vskip 1cm
\begin{center}
{\Large \bf Asymmetry of  $\vec p + \vec p \to\gamma+jet+X$ process\\[0.2cm]
and problem of $\Delta G$}
\end{center}

\vskip 1cm
\begin{center}

{\bf G.P.\v {S}koro}\footnote{goran@rudjer.ff.bg.ac.yu},
\vskip 0.3cm
{\it Faculty of Physics, University of Belgrade,\\
Institute of Nuclear Sciences "Vin\v {c}a",\\
Belgrade, Yugoslavia}

\vskip 0.5cm
{\bf
 M.V.Tokarev}\footnote{tokarev@sunhe.jinr.ru}
\vskip 0.3cm
{\it Laboratory of High Energies,\\
Joint Institute for Nuclear Research,\\
141980, Dubna, Moscow Region, Russia}
\end{center}

\vskip 2cm

\begin{abstract}
The $\gamma +jet$ production in $\vec p-\vec p$ collisions at high energies
is studied. Double-spin asymmetry $A_{LL}$ of the process is calculated
by using Monte Carlo code SPHINX. The predictions for $A_{LL}$ of the 
$\gamma +jet$ production in the $\vec p-\vec p$  collisions 
using the spin-dependent gluon distributions with positive and negative 
sign of $\Delta G(x,Q^2)$  are made at RHIC energies.
A possibility to extract the spin-dependent gluon  distribution 
$\Delta G(x)$ at STAR experiment
and to discriminate between different theoretical scenarios is discussed.
\end{abstract}

\newpage

{\section{Introduction}}          

\vskip 0.3cm

  The  accelerator complex of polarized protons (RHIC-spin) at Brookhaven National
Laboratory  will be first proton-proton collider accelerating
proton beam up to energy of $250~GeV$ with polarization $\simeq 70\%$
and luminosity up to $L \simeq 2\cdot10^{32}~cm^{-2}s^{-1}$ at
$\sqrt s = 500~GeV$. RHIC-spin will have unique capability to
 study polarization phenomena such as direct $\gamma$, jet, dijet,
 Drell-Yan lepton
pairs and $W^{\pm}, Z^0$ \cite{RHIC,Bunce,Bunce1}
production and to perform the test of perturbative and non-perturbative
QCD. The  main goal is to obtain direct information on 
the spin-dependent gluon and sea quark 
distribution and to clarify  nature of proton spin.

In our previous works \cite{NC1,NC2}, we studied the double-spin asymmetry
$A_{LL}$  of jet, dijet 
and direct-$\gamma$ production
in  $\vec p-\vec p$  collisions at RHIC energies 
as a function of sign and shape of spin-dependent gluon
distribution $\Delta G$ and
made predictions of $A_{LL}$
for the STAR experiment planed at RHIC \cite{Bunce1,Bland}.
We would like to note that global fits \cite{QCDfits} to the present inclusive
DIS data cannot even fix the sign of $\Delta G$, not mentioning
its magnitude \cite{Cheng}. For example, 
one of the analysis \cite{Forte} favors a
positive sign, while the recent NLO analysis \cite{Ghosh}
yields negative first moment of the gluon density. 
Both positive and negative values of
the sign  of  $\Delta G(x, Q^2)$ over a wide kinematic  range
($10^{-3}<x<1$) were considered in \cite{tok1}.
The possibility to draw conclusions on the sign of the
spin-dependent gluon distribution, $\Delta G(x, Q^2)$, from existing
polarized DIS data have been studied in \cite{tok2}.
The result of the DIS data analysis \cite{tok2}
on $g_1^n$ supports the conclusion  that the sign
of $\Delta G(x,Q^2)$ should be positive. 
Also, the recent HERMES result \cite{Hermes} indicates
that $\Delta G/G$ is positive in the intermediate $x$ region.
Nevertheless, the additional confirmations on sign of $\Delta G$
are required and the experiments for direct measuring of the $\Delta G/G$
are necessary.
We also consider that the NLO QCD analysis  with negative sign of 
$\Delta G(x,Q^2)$ of DIS data on $g_1$  structure function  could give 
additional constraints of quark and gluon distributions moreover 
usually used \cite{QCDfits}.

The measurement of the $\gamma +jet$ production
in polarized proton-proton collisions
is more perspective for determining $\Delta G(x)$ because the
reaction is dominated by the Compton $qg\rightarrow \gamma q$ scattering
and we can assume that $\Delta q(x)/q(x)$ is known from 
polarized DIS experiments 
\cite{EMC}-\cite{HERM2}. 
Some contributions to the  cross section 
of $\gamma +jet$ production
come from
non-direct mechanisms such as $\bar qq$ annihilation, photon fragmentation
and $k_{\bot}$ smearing \footnote{The discussion on the problem can be found
in \cite{Bunce1} and references therein.}.

The main goal of this work is to made predictions of 
double-spin asymmetry $A_{LL}$ of the 
$\vec {p} + \vec {p} \to \gamma + jet + X$ process and 
to study the possibility of direct
extraction of spin dependent gluon distribution $\Delta G(x)$ 
from this reaction at STAR detector at RHIC.
The determination of ${\Delta}G$ by means of $\gamma +jet$ asymmetry
is connected with large background, similar as in the case of
prompt photon production. Two closely spaced photons resulting
from decay of high $p_T$ neutral mesons
($\pi^0$, $\eta$) can produce fake 'direct photon signal' in detector.
Because the cross-sections of the main subprocesses
for 'direct' mesons production in the $pp$ collision ($qq\rightarrow qq$,
$qg\rightarrow qg$,  $gg\rightarrow gg$) are much higher than the
cross-section for Compton scattering, we can expect the large
yield of background (direct meson + jet) events.
So, the additional aim of the present paper is to 
study the kinematic properties of signal ($S$) and background ($B$)
and to find the corresponding algorithm for background reduction.

The paper is organized as follows. The different sets of 
spin-dependent parton distributions used in Monte Carlo (MC) simulations
are described in Section 2. The  procedure for calculation of $A_{LL}$
and extraction of ${\Delta}G(x)$, together with the results of MC simulations
at $\sqrt s=200~GeV$ are given in Section 3. 
Conclusions are summarized  in Section~4.

\vskip 0.5cm

{\section{Spin-dependent gluon distribution}}    

\vskip 0.3cm

The 3 sets of spin-dependent parton distributions \cite{tok1}
and \cite{Altarelli} are used to calculate the $\gamma +jet$ 
 asymmetry $A_{LL}$.
First  set is based  on work of Altarelli and Stirling \cite{Altarelli}
and include a scenario with large gluon polarization $\Delta G$.
Second and third ones have been obtained by the phenomenological
method \cite{tok1}
including some constraints on the signs of valence and
sea quark distributions, taking  into account the axial
gluon anomaly and utilizes results on integral quark contributions
to the nucleon spin.  Based on the analysis of experimental data on
deep inelastic  structure function $g_1$
the parameterization of
spin-dependent parton distributions for both
positive  and negative sign of  $\Delta G$ have been constructed.
It was found that the $Q^2$ evolution of structure function $g_1$
is sensitive to sign of $\Delta G$.
We would like to note the both sets of distributions describe
experimental data very reasonable.
We shall denote  $\Delta G^{>0}$  and   $\Delta G^{<0}$
sets of spin-dependent parton distributions obtained
in \cite{tok1}  with positive and negative
sign of $\Delta G$, respectively.
It was shown in \cite{tok3} that  phenomenological method
reproduces the main features of the NLO QCD $Q^2$-evolution
of proton, deuteron  and neutron  structure function $g_1$.
Therefore the constructed spin-dependent quark and gluon distributions
can be reasonably used to study the  asymmetry 
$A_{LL}$
 of $\gamma +jet$ production
in $\vec p-\vec p$ collisions too.

Figure 1(a) shows the dependence  of the ratio $\Delta G(x,Q^2)/G(x,Q^2)$
on $x$ at $Q^2=50\ (GeV)^2$ for gluon distributions
\cite{Altarelli} and   $\Delta G^{>0}$ \cite{tok1}.  The curve
\cite{Altarelli} increases with $x$  and then decreases to zero.
The behavior is due to the structure of the ratio
$\Delta G/G\sim x^{\alpha}\dot (1-x)^{\beta}$.
The monotonous increase of an second curve with $x$ connects
to other asymptotic  of the ratio
$\Delta G/G\sim x^{\delta}$  at $x \rightarrow 1$.
Figure 1(b) shows the dependence  of the ratio $\Delta G(x,Q^2)/G(x,Q^2)$
on $x$ at $Q^2=50\ (GeV)^2$ for gluon distributions
$\Delta G^{>0}$ and $\Delta G^{<0}$ \cite{tok1}.

The integral contributions  of gluons  
$\Delta g(Q^2)=\int_0^1 \Delta G(x,Q^2) dx$ to the
proton's spin   are found to be $\Delta g^{>0} \simeq 2$, 
$\Delta g^{<0} \simeq -3.4$ at  $Q^2=10\ (GeV)^2$ \cite{tok1}, and
$\Delta g \simeq 3$ at  $Q^2=10\ (GeV)^2$ \cite{Altarelli}.
Because integral quark  contributions to the proton's spin
is practically the same for all the parameterizations,
the difference in asymmetry  $A_{LL}$ should reflects the
difference in shape of gluon distribution (and sign of  $\Delta G$) used.
We would like to note that large value of gluon polarization corresponds 
to relatively small contribution of $q\bar{q}\rightarrow g\gamma$  
process to  the  final $\gamma + jet$ asymmetry. 
Because this process has the same final state as a signal, it is  
the background for determining the $\Delta G$.
But, our calculations
shows that, at RHIC energies, the cross-section for 
$q\bar{q}\rightarrow g\gamma$ process is
practically for an order of magnitude smaller than the cross-section for
$qg\rightarrow \gamma q$ process, for all spin-dependent PDF's used, so
such a kind of  background could be neglected, especially if we compare it
with direct mesons yields.

{\section{ Monte Carlo simulation results}}

\vskip 0.3cm

For detailed study of the $\vec {p} + \vec {p} \to \gamma + jet + X$ process 
 we have used the Monte Carlo code SPHINX
\cite{SPHINX} which is 'polarized' version of  PYTHIA \cite{PYTHIA}.
Jet reconstruction was done by the
JETSET-subroutine LUCELL \cite{PYTHIA}. This routine defines
jets in the two-dimensional (${\eta},{\phi}$)-plane,
$\eta$ being pseudorapidity and $\phi$ the azimuthal angle.
STAR detector covers
full space in azimuth and pseudorapidity region $-1<{\eta}<2$.

In order to have segmentation expected at STAR
(${\Delta}{\eta}{\times}{\Delta}{\phi}=
{0.1\times}0.1$)
we used 30 $\eta$-bins and 60 $\phi$-bins
in our calculation procedure.
The values of LUCELL-subroutine parameters
$E_{\bot}^{cell}$ and $R=\sqrt{{\Delta}{\eta}^2 +
{\Delta}{\phi}^2}$ were $E_{\bot}^{cell}= 1.5\ GeV$ and $R=0.7$.

The expected resolution of the STAR
Electromagnetic Calorimeter ${\Delta}E/E {\simeq} 0.16/{\sqrt E}$
was also taken into account \cite{LeCompte}.
To obtain total rates ($Rate=\sigma\cdot L$)
of $\gamma + jet$ events at STAR
we taken into account designed integrated luminosity at RHIC:
$L =320~pb^{-1}$ at $\sqrt s = 200~GeV$.
The expected values of polarization of proton beams at RHIC 
($P_{b1}=P_{b2}=P=0.7$)
were also used in the analysis.

All of the SPHINX calculations were done with a cut on the partonic
transverse momentum $p_T > 10~GeV/c$. The simulations include initial-state
radiation, final-state radiation, fragmentation and particle decay effects.

\vskip 0.5cm

{\subsection{Identification of $\gamma + jet$ events and background reduction}}    

\vskip 0.3cm

After generating the signal
event sample
we employ the following cuts for reconstruction of candidate $\gamma + jet$
event:
\begin {itemize}
\item there is exactly $1$ reconstructed jet  with $-0.3<{\eta_{jet}}<1.3$,
$E_T^{jet} >10~GeV$
\item and there is a $1$ photon satisfying $-1<{\eta_{\gamma}}<2$,
$p_T^{\gamma} > 10~GeV/c$. 
\end{itemize}

We would like to note that study of the background enter at this level
of the analysis.
The direct meson background was simulated by using all possible $2\to 2$
partonic hard scattering processes. 
The event is treated as a candidate direct photon event if there is a  
high $p_T$ ($p_T > 10~GeV/c$) neutral meson ($\pi^0 , \eta$) in 
coincidence with the $1$
reconstructed jet ($E_T^{jet} >10~GeV$) satisfying the pseudorapidity cuts defined above. 

Figure 2 shows the comparison of the $direct~meson +jet$ to
$direct~photon +jet$ yields at STAR in $\vec p - \vec p$ collisions at 
$\sqrt s=200~GeV$ as a function of transverse momentum $p_T$
for the spin-dependent parton distribution $\Delta G^{>0}$ \cite{tok1}. 
The signal/background ratio is found to be $S/B \simeq 1/10$.
As seen from Figure 2 it is necessary to impose additional 
cuts for background suppression.

In such cases, the so-called 'isolation cuts' method, 
 developed by UA2 Collaboration \cite{UA2}, 
have been used
for meson background suppression. The candidate direct photon was treated 
as 'isolated' if no
other photons or charged particles fall within a cone of radius $R_{isol}$
around the photon. The idea is that direct meson usually has much
more particles in its 'neighborhood' than direct photon. 

We tried to make more sophisticated algorithm for photon 
isolation comparing the
energy and multiplicities of particles falling within the isolation cone
for different values of $R_{isol}$. The results are shown in Figure 3.
We have used 3 different values of isolation cone radius: $R_{isol}= 0.26$
(UA2 condition), $R_{isol}= 0.5$  and $R_{isol}= 0.7$.

As seen from Figure 3  particles accompanying the
direct photon are softer 
than those for the direct meson.
Also, the use of larger cone radius allow better
reduction of background without significant loss of signal. 
So, the conclusion is that the combination of cuts on multiplicity and
energy will give better background suppression.
Our results show that highest gamma efficiency $P_{\gamma/\gamma}$
  and meson rejection $1-P_{\gamma/\pi^0}$ 
(here $P_{\gamma/\pi^0}$ is meson contamination) values are obtained
under following conditions: the isolation cone radius is $R_{isol}= 0.7$ and 
summed energy of particles (if there are any) around a direct photon 
candidate must be lower than $1~GeV$.  In such a case, $90\%$ of mesons
were rejected with gamma efficiency $P_{\gamma/\gamma}$ of $88\%$.

Figure 4 shows comparison of the $direct~meson + jet$ to
$direct~photon +jet$ yields at STAR in $\vec p - \vec p$ collisions at 
$\sqrt s=200~GeV$  as a function of transverse momentum $p_T$   
for spin-dependent parton distribution $\Delta G^{>0}$ \cite{tok1} 
with the isolation cuts imposed. 

Additional event-by-event discrimination between single photons and
photon pairs (from neutral mesons decays) at STAR is based on measurement of 
transverse profile of the electromagnetic shower in Shower Maximum Detector
(SMD). It is expected that this approach will rejects $\simeq 80\%$ of
photon pairs from meson decays with single gamma
efficiency of $\simeq 80\%$ \cite{Vigdor, EEMC,BEMC}.

Figure 5 shows comparison of the $direct~meson + jet$ to
$direct~photon +jet$ yields at STAR in $\vec p - \vec p$ collisions at 
$\sqrt s=200~GeV$  as a function of transverse momentum $p_T$   
for spin-dependent parton distribution $\Delta G^{>0}$ \cite{tok1}, 
after the additional 
$meson/\gamma$ discrimination with the help of
Shower Maximum Detector.

Finally, we can conclude that only combination of isolation cuts and
$meson/\gamma$ discrimination with the help of SMD will be able to 
reduce the background in the whole $p_T$ region of interest for
measuring of 
double-spin asymmetry $A_{LL}$ 
and extracting of $\Delta G(x)$ at STAR.

\vskip 0.3cm

{\subsection{Asymmetry of $\gamma+jet$ production}}

\vskip 0.3cm    

In Monte Carlo simulations, the double spin
asymmetry is defined through the difference of cross-sections
(numbers of $\gamma + jet$ events) for antiparallel 
(${\uparrow}{\downarrow}$)
and parallel (${\uparrow}{\uparrow}$) spins of colliding protons:

\begin{equation}
A_{LL}=\frac{\sigma^{{\uparrow}{\downarrow}} -
\sigma^{{\uparrow}{\uparrow}}}{\sigma^{{\uparrow}{\downarrow}}
+ \sigma^{{\uparrow}{\uparrow}}} =\frac{1}{P_{b1}P_{b2}}
 \frac{N_{\gamma +jet}^{{\uparrow}{\downarrow}} -
N_{\gamma + jet}^{{\uparrow}{\uparrow}}}
{N_{\gamma + jet}^{{\uparrow}{\downarrow}}
+ N_{\gamma + jet}^{{\uparrow}{\uparrow}}},
\end{equation}

with the statistical  error:

\begin{equation}
{\delta}A_{LL}= {\frac {1} {P_{b1}P_{b2 }}}    \sqrt{\frac{1-(P_{b1}P_{b2}A_{LL})^2}
{N_{\gamma + jet}^{{\uparrow}{\downarrow}}
+ N_{\gamma +jet}^{{\uparrow}{\uparrow}}}},
\end{equation}
where $P_{b1}$ and $P_{b2}$ are values of the polarization of proton beams.
In our case, only the $\gamma +jet$ events that passed all the above cuts
 were used for calculation of $A_{LL}$ and extraction of $\Delta G(x)$.

Figure 6 shows the asymmetry $A_{LL}$   of $\gamma + jet$ production  
in polarized $pp$ collisions at $\sqrt s= 200$~GeV 
for 3 different sets of spin-dependent PDF's
($\Delta G^{>0}$ \cite{tok1}, $\Delta G^{<0}$ \cite{tok1} 
and \cite{Altarelli}) 
as a
function of transverse momentum $p_T$. The absolute value of $A_{LL}$ increases
with $p_T$ for all spin-dependent PDF's used. There is  a clear difference
in $A_{LL}$ between positive and negative sign of $\Delta G$ as well as between
PDF's $\Delta G^{>0}$ \cite{tok1} and \cite{Altarelli} in the region
$p_T =(10-25)~GeV/c$.

\vskip 0.3cm

{\subsection{Partonic kinematics reconstruction}}

\vskip 0.3cm

The Bjorken $x$-values of colliding partons is reconstructed by means of
measured $\gamma$ and jet transverse momenta $p_T$ and pseudorapidities:
\begin{equation}
x_1 = \frac{2p_T}{\sqrt {s}} \frac{(e^{\eta_{\gamma}}+e^{\eta_{jet}})}{2},
\end{equation}
\begin{equation}
x_2 = \frac{2p_T}{\sqrt {s}} \frac{(e^{-\eta_{\gamma}}+e^{-\eta_{jet}})}{2}.
\end{equation}
The usual procedure \cite{Vigdor, sphinx1} is to associate the
larger of the $x_1,x_2$ values with the quark  
\begin{equation}
x_q = max[x_1,x_2],
\end{equation}
and the
smaller one with the gluon momentum fraction
\begin{equation}
x_g = min[x_1,x_2],
\end{equation}
with additional condition:
\begin{equation}
max[x_1,x_2]>0.2.
\end{equation}

Comparison of the simulated to reconstructed values of the initial-state
parton kinematics for the events that passed all above cuts is shown in
Figure 7.

\vskip 0.3cm

{\subsection{Extraction of $\Delta G(x)$}}       

\vskip 0.3cm

The main formula for extraction of $\Delta G(x)$ from $\gamma + jet$ 
coincidence events can be written in the following form
\cite{Vigdor}:
\begin{equation}
\Delta G(x_g)=\frac{1}{P_{b1}P_{b2}} 
\frac{N_{x_g}^{{\uparrow}{\downarrow}}-N_{x_g}^{{\uparrow}{\uparrow}}}
{\sum_{i=1}^{N_s} 
{A_{1p}^{DIS}(x_{qi},Q_{i}^2){\cdot}\hat {a}_{LL}(\cos {\theta_i})}/
{G(x_{gi},Q_{i}^2)}},
\end{equation}
where $N_{x_g}^{{\uparrow}{\downarrow}}$ and 
$N_{x_g}^{{\uparrow}{\uparrow}}$ are number of $\gamma + jet$ events 
for reconstructed  value of gluon momentum fraction $x_g$,
for antiparallel 
(${\uparrow}{\downarrow}$)
and parallel (${\uparrow}{\uparrow}$) spins of colliding protons, 
respectively and $N_s = N_{x_g}^{{\uparrow}{\downarrow}} +
N_{x_g}^{{\uparrow}{\uparrow}}$. 
Here $\hat {a}_{LL}(\cos {\theta_i})$ is 
the partonic spin-correlation parameter
for $qg\to q\gamma$ process, $\theta_i$ being the scattering angle in
the partonic center-of-mass (pCM) system. Assuming collinear partonic 
collisions, the absolute value of $\cos {\theta_i}$ can be obtained as:
\begin{equation}
\cos {\theta_i} = \tanh ({\pm \eta_i}),
\end{equation}
where
\begin{equation}
\eta_i = \pm \frac{1}{2}(\eta_{\gamma} - \eta_{jet}).
\end{equation}
The value of the partonic spin-correlation parameter 
$\hat {a}_{LL}(\cos {\theta_i})$ can be calculated in the LO pQCD
\cite{Sof} and for $qg\to q\gamma$ process we have:
\begin{equation}
\hat {a}_{LL} =\frac{{\hat {s}}^2 - {\hat {u}}^2}
{{\hat {s}}^2 + {\hat {u}}^2},
\end{equation}
where $\hat {s}$  and $\hat {u}$ are the Mandelstam variables for the
hard scattering:
\begin{equation}
\hat {s} = x_1 x_2 s,
\end{equation}
\begin{equation}
\hat {u} =-\frac{\hat {s}}{2} (1+\cos {\theta_i}).
\end{equation}

The factor $A_{1p}^{DIS}(x_{qi},Q_{i}^2)$ is a
parameterization of the $A_1$ structure function for the proton obtained
by fitting the results of polarized DIS experiments. The value
 $x_{qi}$ is the
quark momentum fraction and $Q_{i}^2$ is momentum transfer scale, 
assumed to be 
$Q_{i}^2 = p_{T}^{2}/2$. For the calculations done here, we used the
following parameterization of $A_{1p}^{DIS}$ \cite{Savin}:
\begin{equation} 
A_{1p}^{DIS}(x_{qi},Q_{i}^2){\sim}A_{1p}^{DIS}(x_{qi})
=0.01902 + x_{qi}^{-0.01163}(1-e^{-1.845x_{qi}}).
\end{equation}
The parameterization of unpolarized gluon structure function
$G(x_{gi},Q_{i}^2)$,
evaluated at reconstructed value of $x_g$ and evolved to the
value of $Q_{i}^2$,  was taken from \cite{Gl}.

The results of extraction procedure for spin-dependent parton distribution $\Delta G^{>0}$ \cite{tok1} 
at colliding energy of $\sqrt s=200~GeV$ is shown in Figure 8.
In order to discuss the quality of extraction procedure, we show in Figure 8
the comparison  of the input polarized gluon distribution to values
extracted from the simulated data as a function of the reconstructed
$x_{gluon}$. The input polarized gluon distribution is evolved to the most
probable value $Q^2 = 50~(GeV/c)^2$. 

In general, theoretical input and the
values extracted from the data agree very well. 
The agreement with results obtained in \cite{Bland} is also reasonable. 
The  systematic
discrepancy for $x_{gluon} < 0.08$ reflects the ambiguity in assigning
$x_{1(2)}$ values to gluon (quark) in the small $x$ region and ambiguity in
the choice of the sign for $\cos {\theta_i}$.
Also, the behavior of extracted polarized gluon distribution at 
higher $x$-values reflects the role of condition (7) and results 
in relatively small change of the shape of the distribution 
for $x_{gluon} > 0.2$. 
Nevertheless, this result shows
that $\Delta G(x)$ can be extracted at STAR from the measurements of
$\gamma + jet$ production in $\vec p-\vec p$ collisions. At $\sqrt s=200~GeV$,
STAR detector will give us the full information about $\Delta G(x)$ in
the region $x_{gluon} = (0.03-0.3)$.

Figure 9 shows reconstructed values of $x\Delta G(x)$ versus 
reconstructed values of $x_{gluon}$
for 3 different sets of spin-dependent PDF's
($\Delta G^{>0}$ \cite{tok1}, $\Delta G^{<0}$ \cite{tok1} 
and \cite{Altarelli}). One can see that measurements at STAR will be able
to discriminate between scenarios with positive and negative sign of
$\Delta G$. In the case of positive sign of $\Delta G$, extraction of
$\Delta G(x)$ at $\sqrt s=200~GeV$ will give us possibility to 
distinguish between theoretical spin-dependent 
PDF's $\Delta G^{>0}$ \cite{tok1} and
\cite{Altarelli} in the region $x_{gluon} = (0.03-0.2)$.

\vskip 0.5cm

{\section{Conclusions}}  
           
The $\gamma +jet$ production in $\vec p-\vec p$ collisions at RHIC energies
was studied. Double-spin asymmetry $A_{LL}$ of the process is calculated
by using Monte Carlo code SPHINX taking into account parameters 
of the STAR detector.
Special attention was paid on the procedure for suppression of meson
background. The  algorithm for isolation of direct gamma based on 
'isolation cuts' method was developed.  
A procedure for extraction of $\Delta G(x)$ from simulated data
sample was described in detail.
The obtained results show that highest gamma efficiency and meson rejection  
values are obtained under following conditions:
 the isolation cone radius is $R_{isol}= 0.7$ and 
summed energy of particles (if there are any) around a direct photon 
candidate must be lower than $1~GeV$.  In such a case, $90\%$ of mesons
were rejected with gamma efficiency of $88\%$. 
Also, we concluded that only combination of isolation cuts and
$meson/\gamma$ discrimination with the help of Shower Maximum Detector
will be able to  reduce the background in the whole $p_T$ region 
of interest for measuring of double-spin asymmetry $A_{LL}$ 
and extracting of the spin-dependent gluon  distribution $\Delta G(x)$ 
in the region $x_{gluon} = (0.03,0.3)$  at $\sqrt s =200~GeV$ at STAR.

Thus, the obtained results allow us to hope that direct experimental
information on a   sign of the  spin-dependent gluon distribution
will give constraints  on the quark and gluon distributions and to
clarify a role of quark and gluon orbital momenta in proton's spin
composition \cite{Jaffe}.

\vskip 0.5cm


\vskip 0.5cm

\newpage

\begin{minipage}{4cm}

\end{minipage}

\vskip 2cm

\begin{center}

\hspace*{0.5cm}

\parbox{6.5cm}{\epsfxsize=6.5cm \epsfysize=6.5cm \epsfbox[5 5 500 500]
{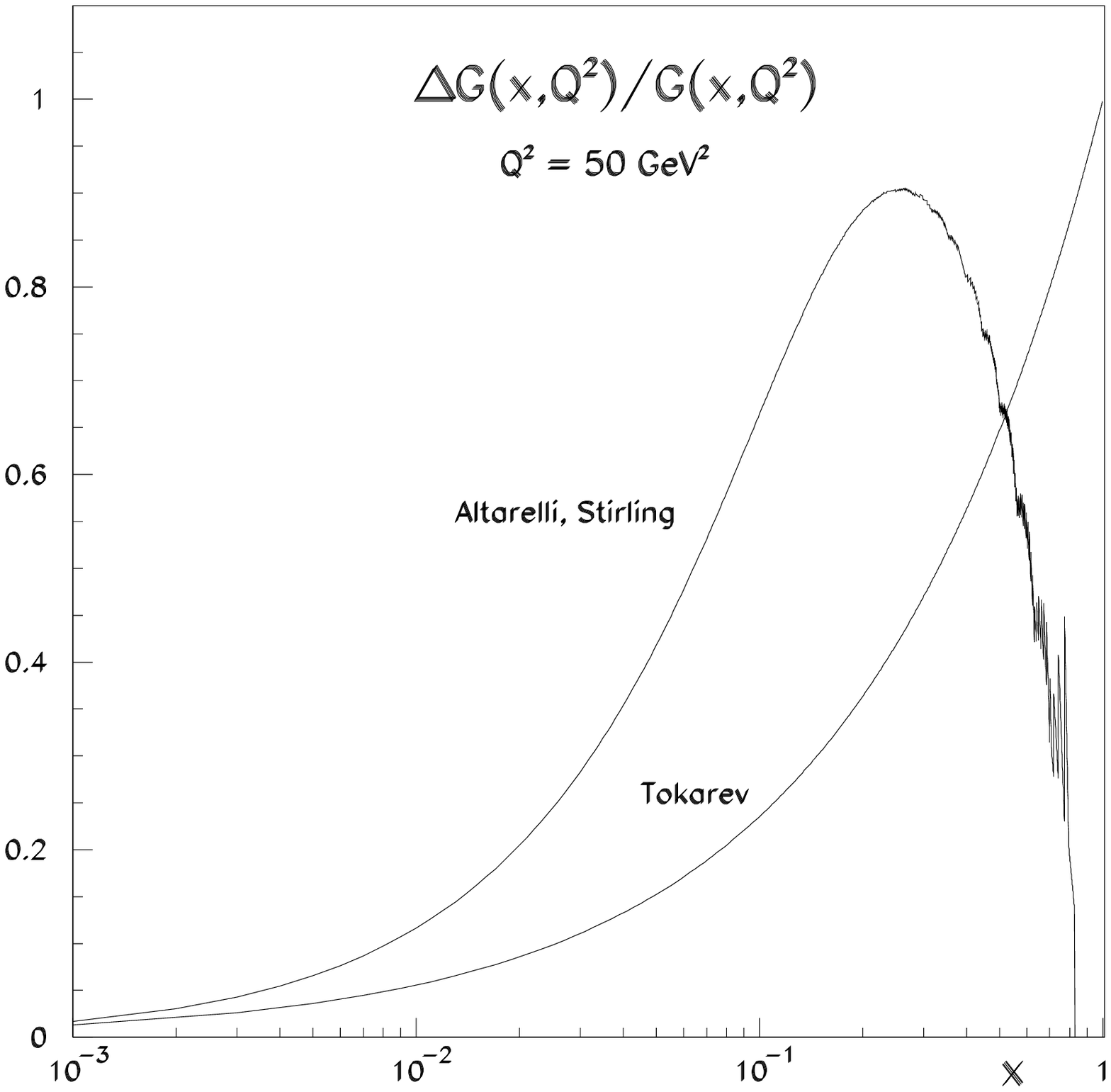}{}}

\vskip -0.5cm

\hspace*{1. cm} a) \\

\end{center}

\vskip 1.cm

\begin{center}

\hspace*{0.5cm}

\vskip 1.5cm

\parbox{6.5cm}{\epsfxsize=6.5cm \epsfysize=6.5cm \epsfbox[5 5 500 500]
{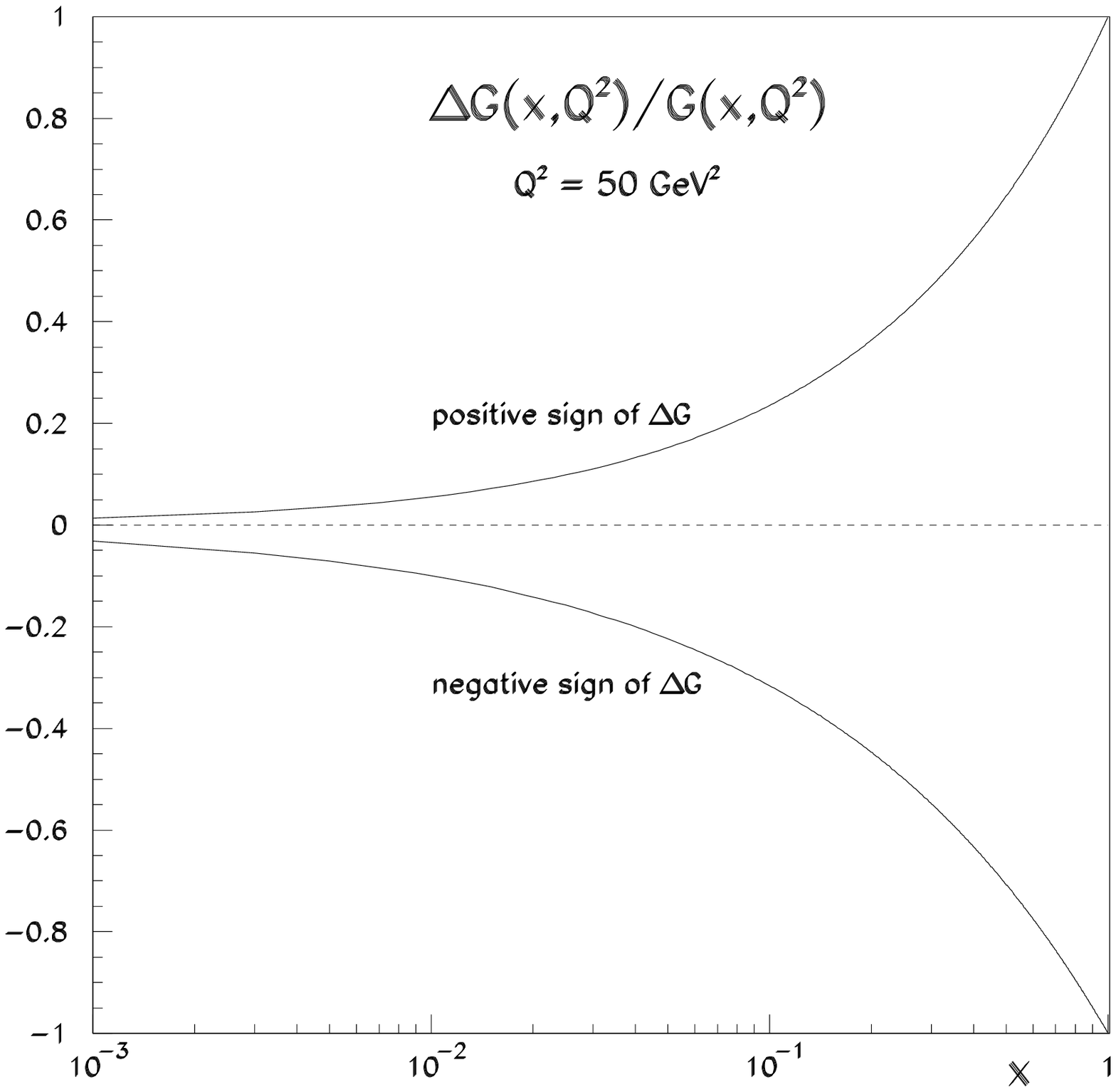}{}}

\vskip -1.5cm

\hspace*{1. cm}  b)\\

\end{center}

\vskip -0.5cm

{\bf Figure 1}

The ratio $\Delta G(x,Q^2)/G(x,Q^2)$ of  polarized and unpolarized
gluon distributions as a function of $x$
for two different parameterizations
(a) 
$\Delta G^{>0}$ \cite{tok1},
  \cite{Altarelli} 
and
(b)
$\Delta G^{>0}$ \cite{tok1},
$\Delta G^{<0}$ \cite{tok1} 
   at $Q^2$=50 GeV$^2$.



\newpage

\begin{minipage}{4cm}
\vspace*{4cm}
\end{minipage}
\vskip 3cm
\begin{center}
\parbox{12cm}{\epsfxsize=12cm \epsfysize=12cm \epsfbox[35 5 535 500]
{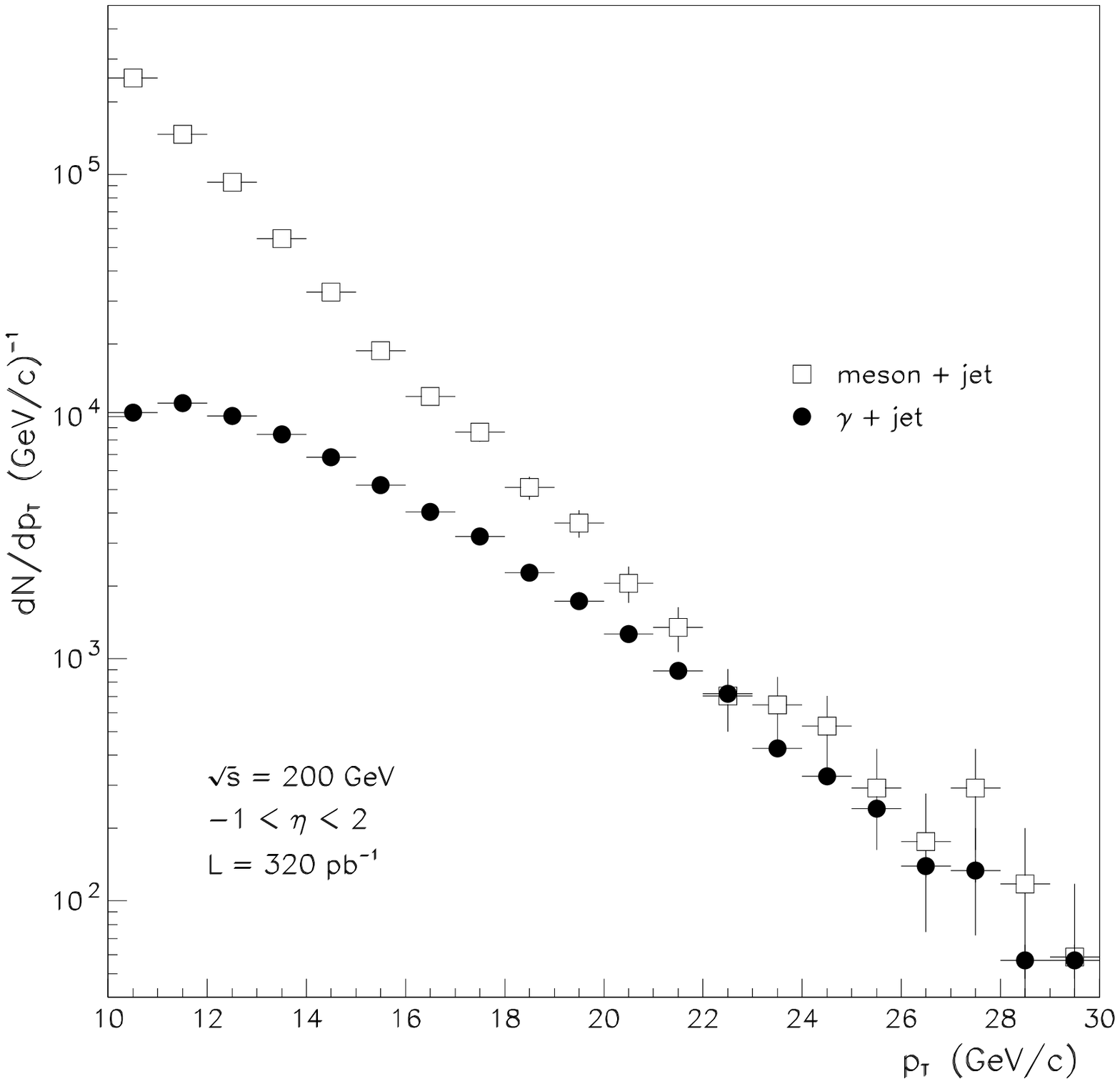}{}}
\end{center}

\vskip -2cm

{\bf Figure 2.} 

Comparison of the $direct~meson +jet$ to
$direct~photon+jet$ yields at STAR in $\vec p - \vec p$ collisions at 
$\sqrt s=200~GeV$ as a function of transverse momentum $p_T$   
for spin-dependent parton distribution $\Delta G^{>0}$ \cite{tok1}.


\newpage

\begin{minipage}{4cm}
\vspace*{4cm}
\end{minipage}
\vskip 3cm
\begin{center}
\parbox{12cm}{\epsfxsize=12cm \epsfysize=12cm \epsfbox[35 5 535 500]
{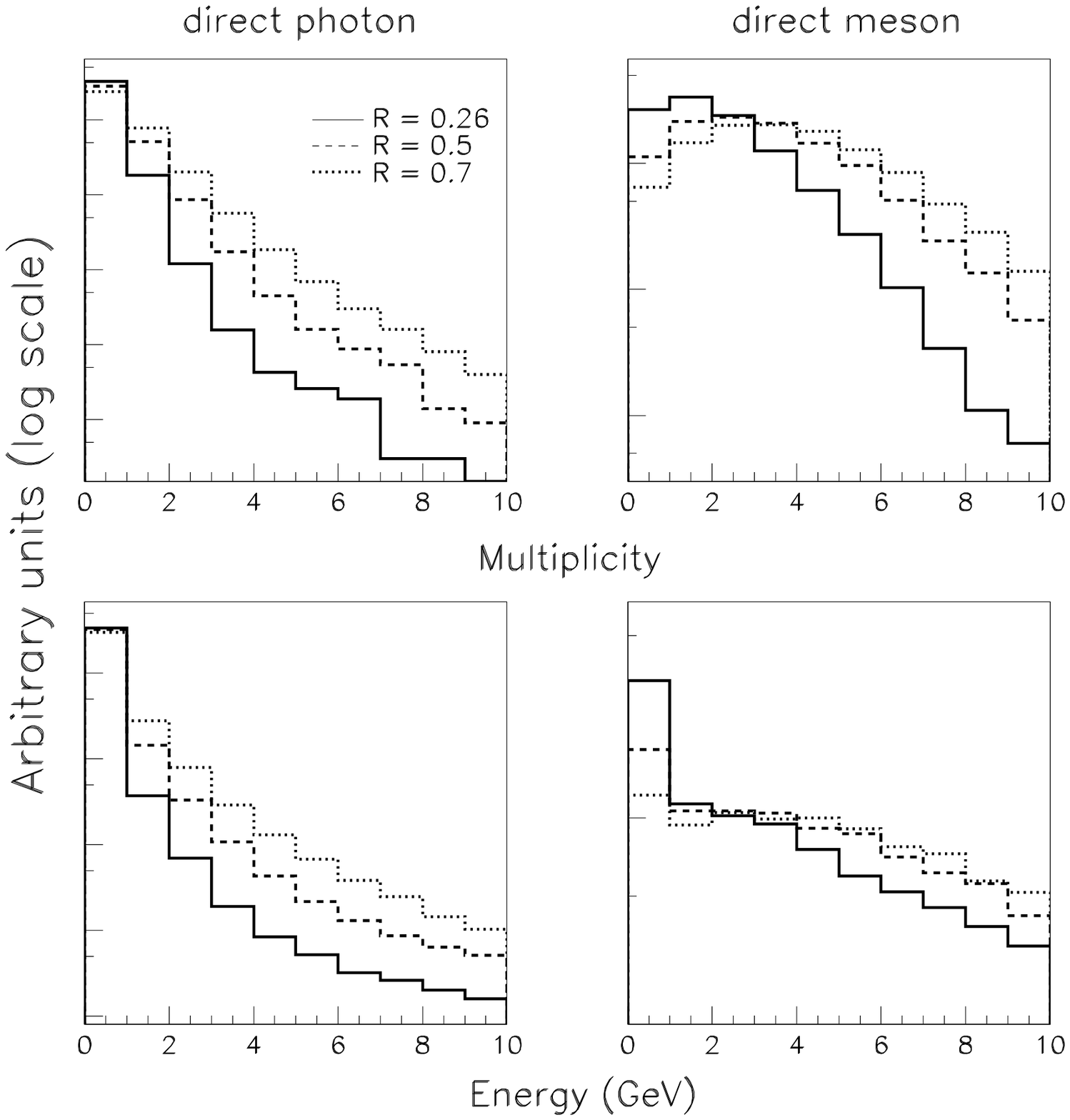}{}}
\end{center}

\vskip -2cm

{\bf Figure 3.}

The multiplicity and summed energy distributions of particles falling
within an isolation cone of radius $R$ around a direct photon (left
column) or a direct meson (right column).


\newpage

\begin{minipage}{4cm}
\vskip 2cm
\end{minipage}
\vskip 2cm
\begin{center}
\parbox{6.5cm}{\epsfxsize=6.5cm \epsfysize=6.5cm \epsfbox[35 5 535 500]
{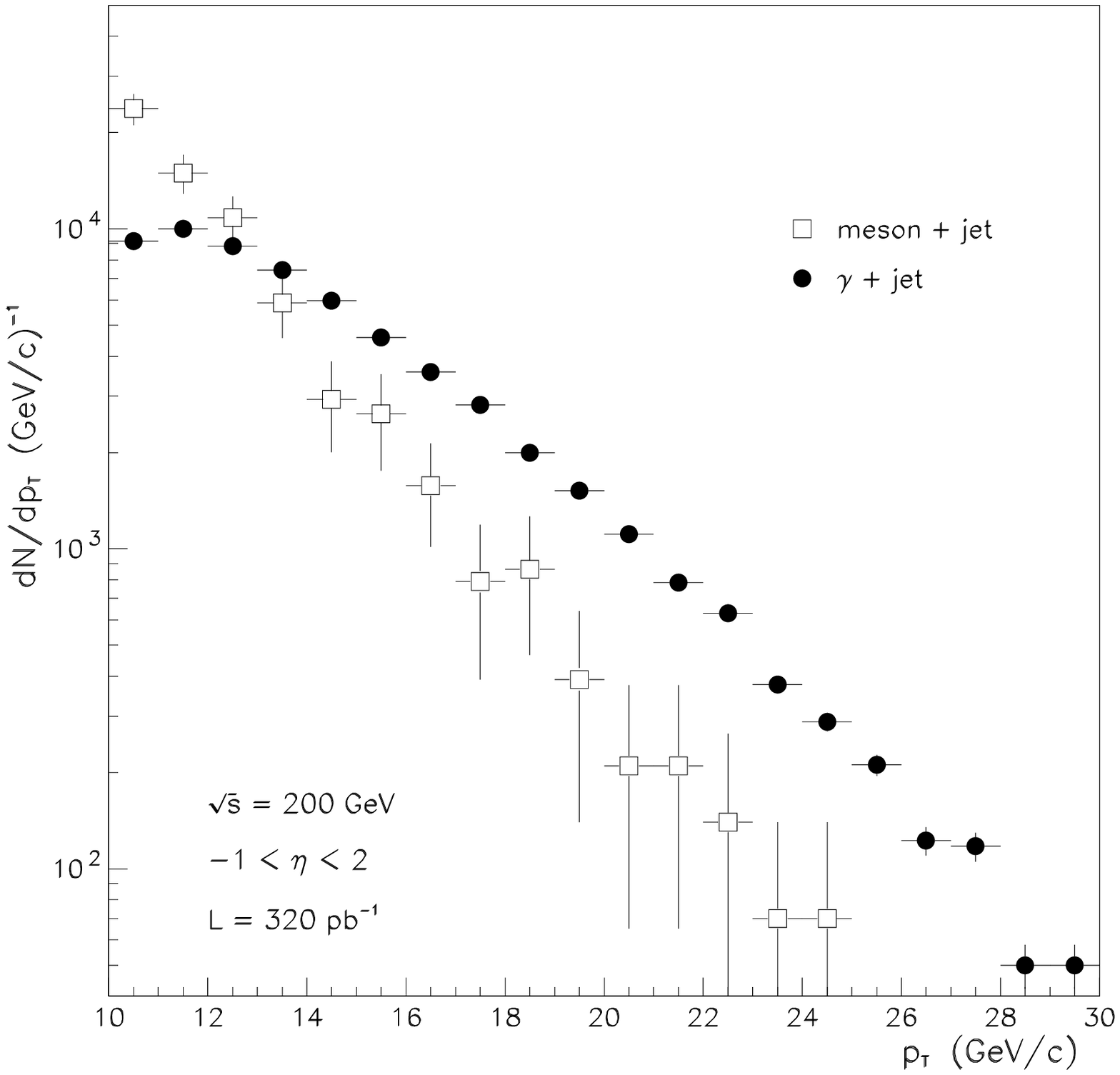}{}}
\end{center}

\vskip -1.5cm

{\bf Figure 4.}

Comparison of the $direct~meson + jet$ to
$direct~photon +jet$ yields at STAR in $\vec p - \vec p$ collisions at 
$\sqrt s=200~GeV$  as a function of transverse momentum $p_T$   
for spin-dependent parton distribution $\Delta G^{>0}$ \cite{tok1}, 
with the isolation cuts imposed (see text).

\vskip 1.5cm

\begin{center}
\vskip 1.5cm
\parbox{6.5cm}{\epsfxsize=6.5cm \epsfysize=6.5cm \epsfbox[5 5 500 500]
{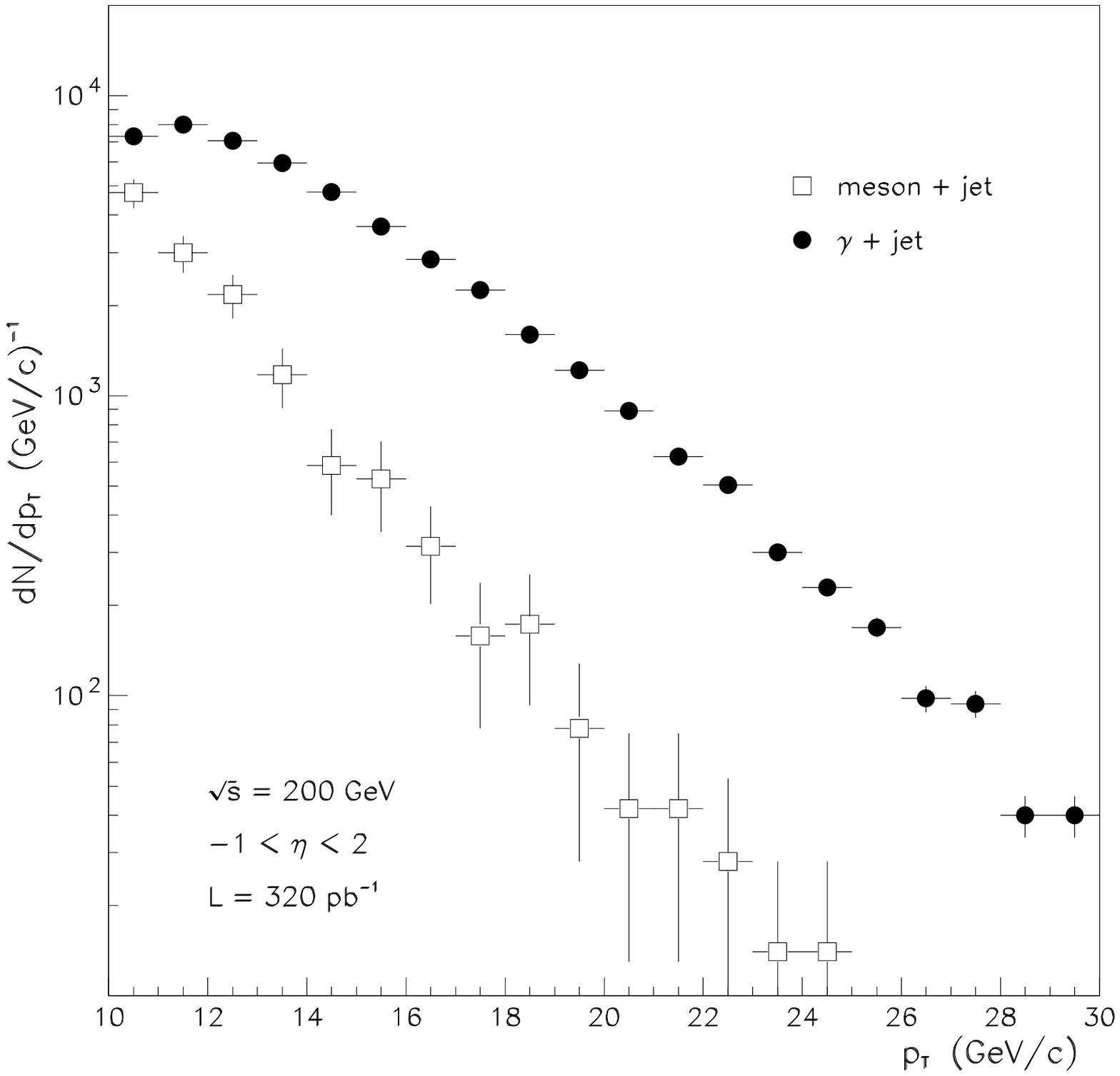}{}}
\vskip -1.5cm
\end{center}

{\bf Figure 5.}

Comparison of the $direct~meson + jet$ to
$direct~photon +jet$ yields at STAR in $\vec p - \vec p$ collisions at 
$\sqrt s=200~GeV$  as a function of transverse momentum $p_T$   
for spin-dependent parton distribution $\Delta G^{>0}$ \cite{tok1}, 
after the additional $meson/\gamma$ discrimination with the help of
Shower Maximum Detector.


\newpage

\begin{minipage}{4cm}
\vskip  2cm
\end{minipage}
\vskip 2cm
\begin{center}
\parbox{6.5cm}{\epsfxsize=6.5cm \epsfysize=6.5cm \epsfbox[35 5 535 500]
{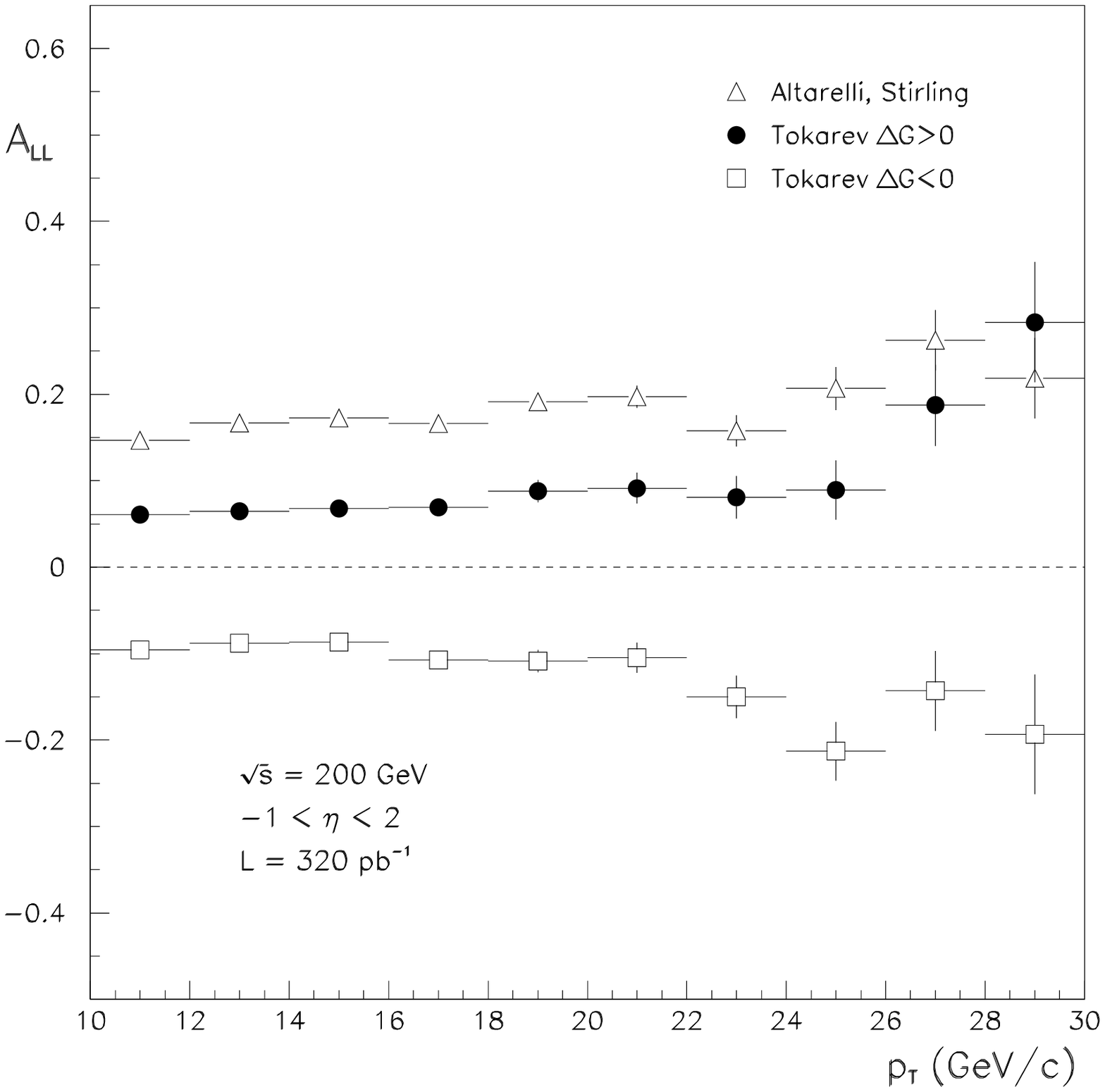}{}}
\end{center}

\vskip -1.5cm

{\bf Figure 6.}

Asymmetry $A_{LL}$ of $\gamma + jet$ production  in polarised $pp$ collisions
at $\sqrt s= 200$~GeV 
for 3 different sets of spin-dependent PDF's
($\Delta G^{>0}$ \cite{tok1}, $\Delta G^{<0}$ \cite{tok1} 
and \cite{Altarelli}) 
as a
function of transverse momentum $p_T$. The errors indicated are
statistical only, based on the expected luminosity of RHIC
and the properties of the STAR detector.

\vskip 3.cm

\begin{center}
\hspace*{-1.cm}

\parbox{6cm}{\epsfxsize=6.cm \epsfysize=6.cm \epsfbox[5 5 500 500]
{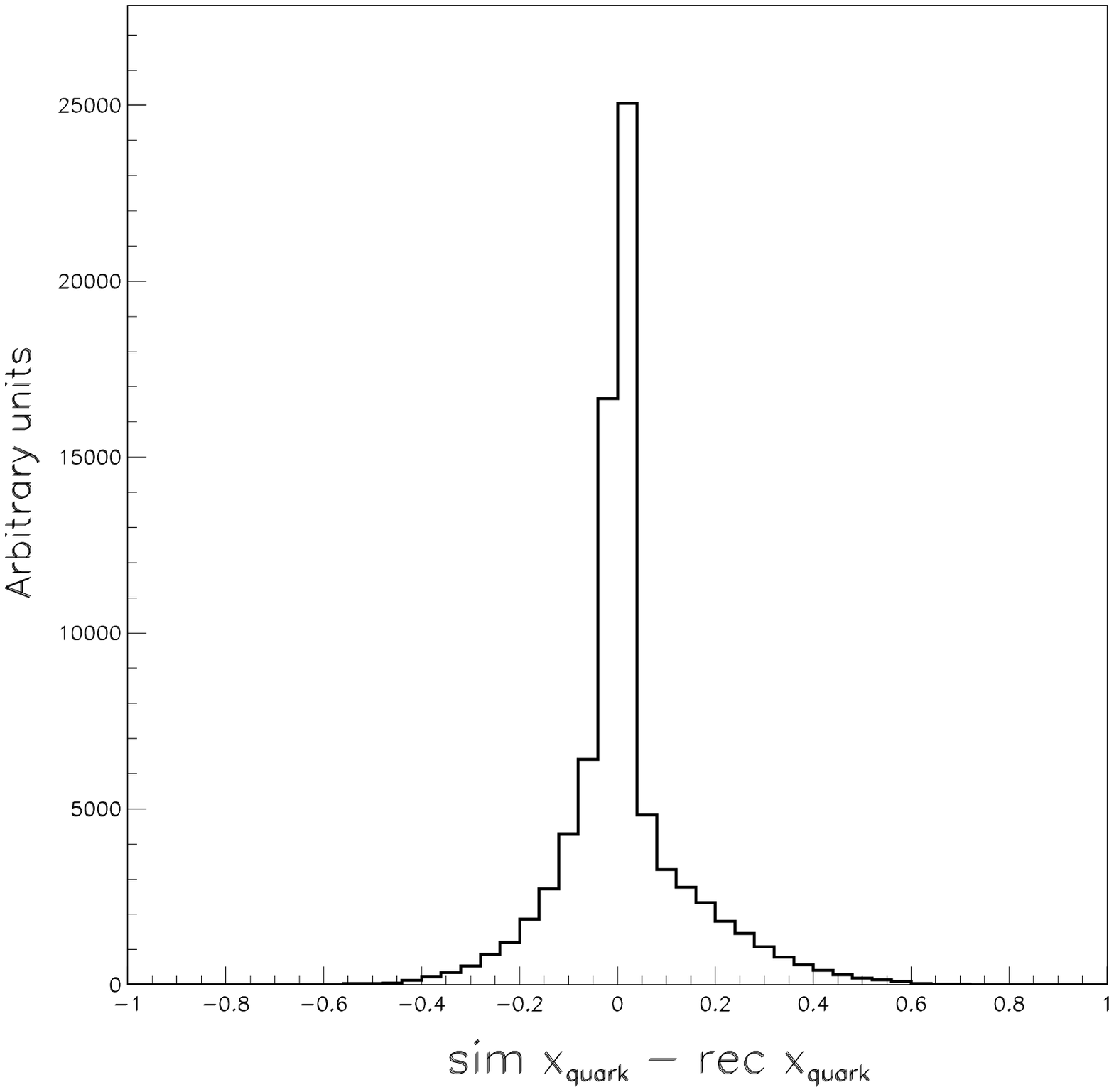}{}}
\hspace*{1cm}
\parbox{6cm}{\epsfxsize=6.cm \epsfysize=6.cm \epsfbox[35 5 535 500]
{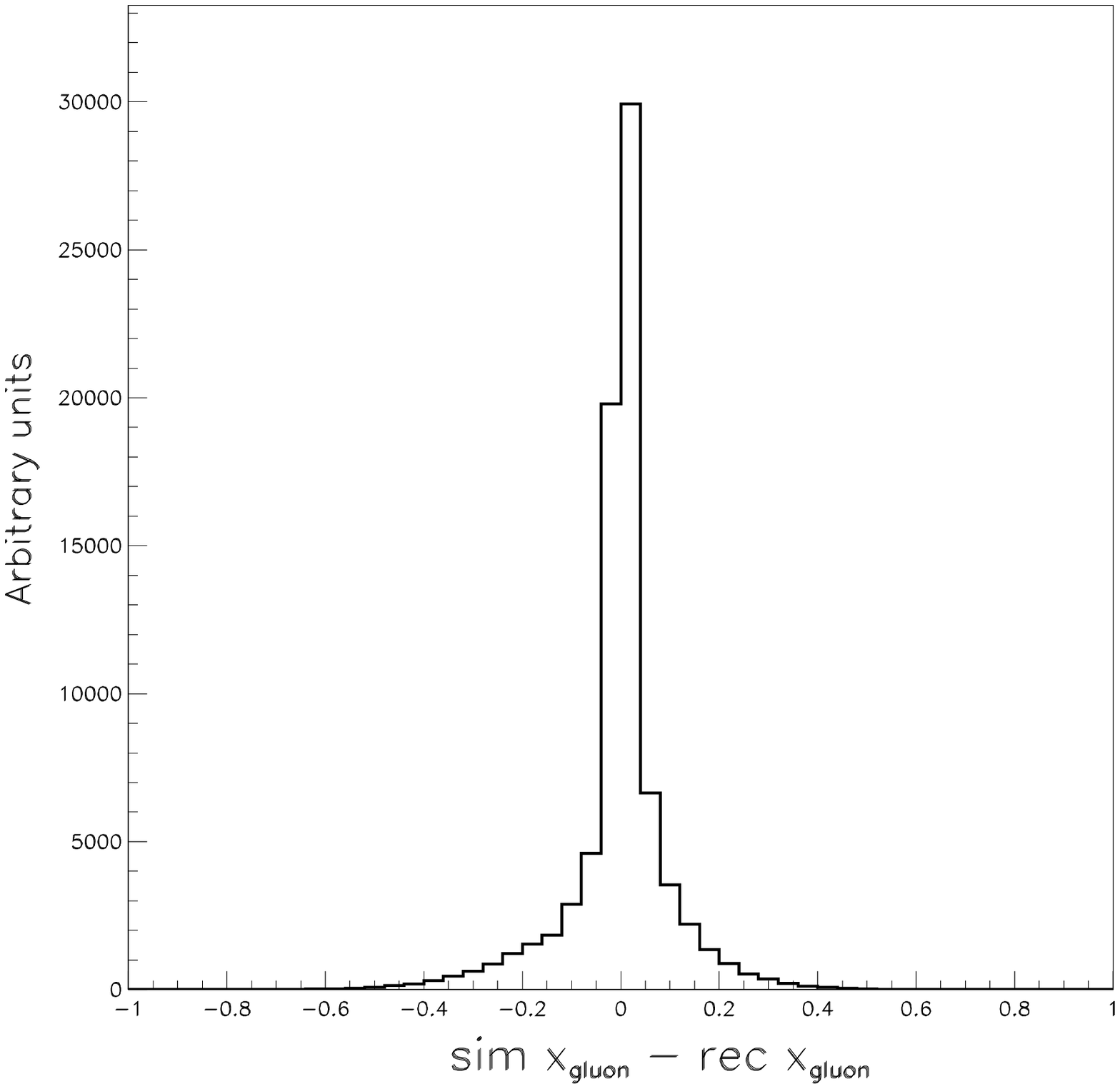}{}}
\vskip -1cm
\hspace*{1. cm} a) \hspace*{6cm} b)\\[0.5cm]
\end{center}

{\bf Figure 7.}
Comparison of the simulated to reconstructed values of the initial-state
parton kinematics.


\newpage

\begin{minipage}{4cm}
\vskip  2cm
\end{minipage}
\vskip 2cm
\begin{center}
\parbox{6.5cm}{\epsfxsize=6.5cm \epsfysize=6.5cm \epsfbox[35 5 535 500]
{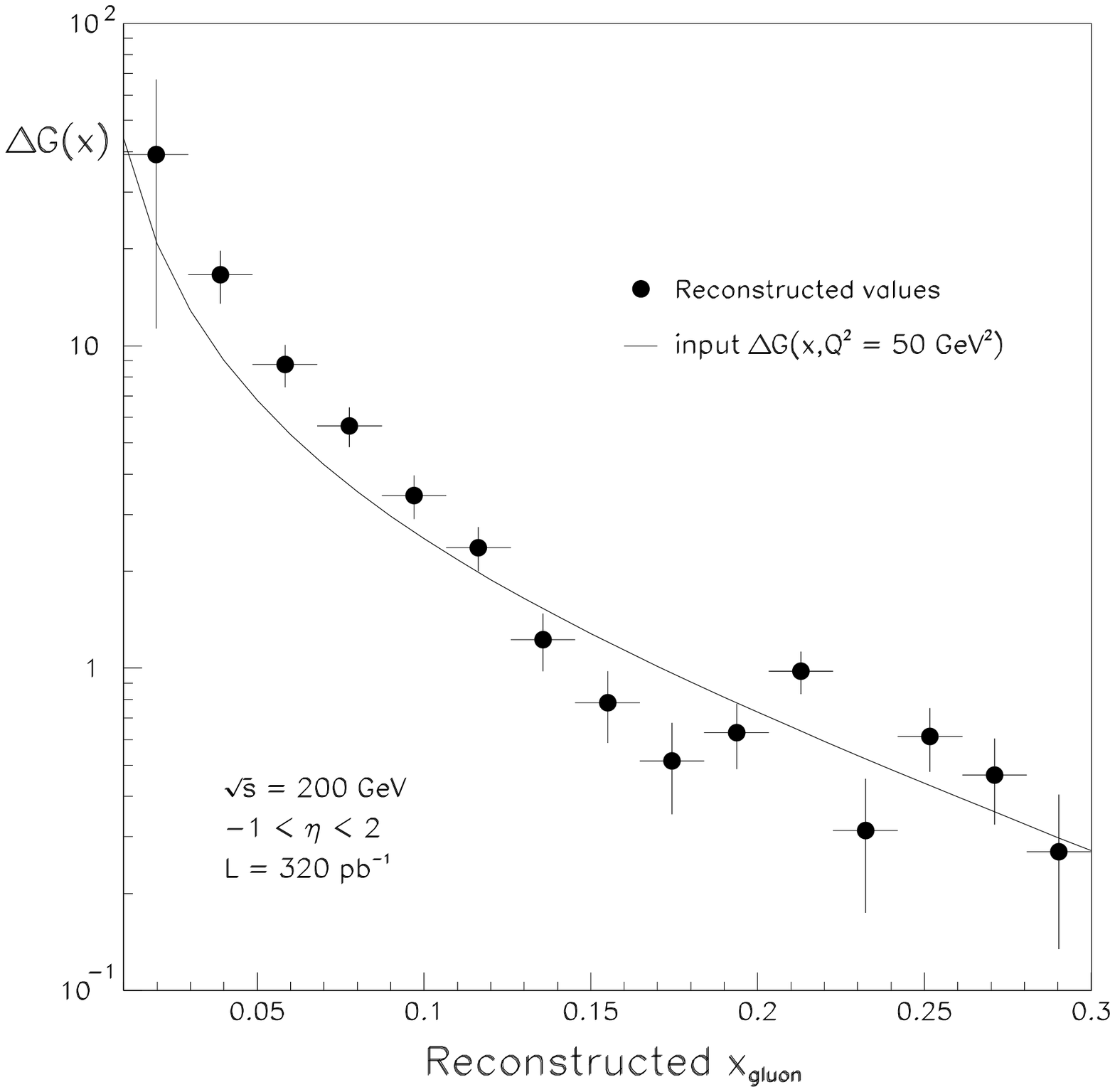}{}}
\end{center}

\vskip -1.5cm

{\bf Figure 8.}

Comparison  of the input polarized gluon distribution (line)
$\Delta G(x)$ to values
extracted from the simulated data (points) as a function of the 
reconstructed $x_{gluon}$. The results are obtained
for spin-dependent parton distribution $\Delta G^{>0}$ \cite{tok1} 
at colliding energy of $\sqrt s=200~GeV$.  

\vskip 1.5cm

\begin{center}
\vskip 1.5cm
\parbox{6.5cm}{\epsfxsize=6.5cm \epsfysize=6.5cm \epsfbox[5 5 500 500]
{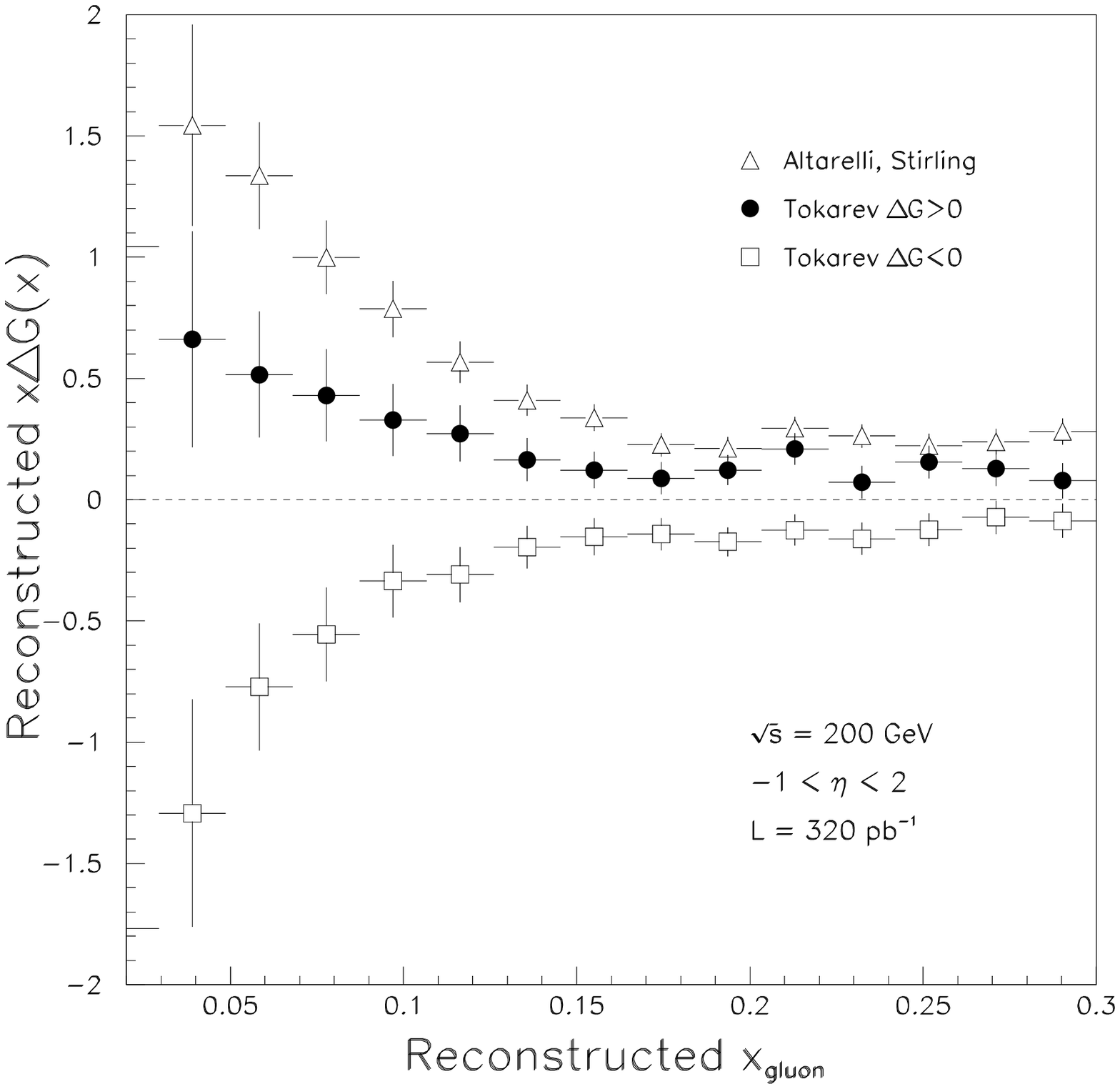}{}}
\vskip -1.5cm
\end{center}

{\bf Figure 9.}

Reconstructed values of $x\Delta G(x)$ versus reconstructed values
of $x_{gluon}$
for 3 different sets of spin-dependent PDF's
($\Delta G^{>0}$ \cite{tok1}, $\Delta G^{<0}$ \cite{tok1} 
and \cite{Altarelli}). 

\end{document}